\documentclass{article}

\usepackage{PRIMEarxiv}

\usepackage[utf8]{inputenc} 
\usepackage[T1]{fontenc}    
\usepackage[hidelinks]{hyperref}       
\usepackage{url}            
\usepackage{booktabs}       
\usepackage{amsfonts}       
\usepackage{nicefrac}       
\usepackage{microtype}      
\usepackage{lipsum}
\usepackage{float} 
\usepackage{fancyhdr}       
\usepackage{xurl}
\usepackage{graphicx}       
\graphicspath{{Graphics/}}     

\title{Examining AI-generated historical narratives and their reception through the example of history POVs on TikTok
}

\author{
  Nina Brolich, Anna Neovesky \\
  University of Erfurt \\
  University of Applied Sciences Erfurt \\
  \texttt{nina.brolich@fh-erfurt.de, anna.neovesky@fh-erfurt.de} \\
}

\begin{document}
\maketitle

\begin{abstract}
This paper examines the history POV trend on TikTok, in which AI-generated 
first-person scenes depict historical events. We use a two-stage empirical approach: an exploratory pilot study and a larger-scale study building up on a dataset obtained through the TikTok Research API. In both studies we analyze the themes of the trend and how the audience responds in the comments. Findings show a dominance of emotionally charged 
contemporary history topics, with historical inaccuracies visible at the caption level.
A comparative comment analysis of Black Death and Holocaust videos, combining manual annotation with DistilBERT-based classification, reveals that topic choice shapes audience response, with Holocaust content attracting disproportionately higher rates of hate speech and disinformation. The paper also reflects on the strengths and limitations of API-based research for studying fast-moving platform trends.
\end{abstract}


\section{Introduction}
The rapid advance of generative AI image and video tools has enabled new forms of AI-generated historical representation, with use cases ranging from retrospectively visualizing women in science\footnote{\url{https://www.uni-bonn.de/de/universitaet/unileben/veranstaltungen/ki-ausstellung-her-mit-den-portr-ai-ts} [Accessed: \today]} to "de-aging" Holocaust survivors for educational campaigns targeting younger audiences\footnote{\url{https://youngagainneveragain.org} [Accessed: \today]}. In January 2026, a coalition of German memorial sites published an open letter demanding that social media platforms take proactive action against so-called \textit{AI slop} – mass-produced AI-generated content – that distorts Holocaust history by trivializing and sensationalizing it.\footnote{\url{https://www.kz-gedenkstaette-dachau.de/nachrichten/offener-brief-konsequentes-vorgehen-gegen-ki-generierte-holocaust-verfaelschungen-auf-social-media-plattformen} [Accessed: \today]}
\begin{figure}
  \centering
  \includegraphics[width=\textwidth]{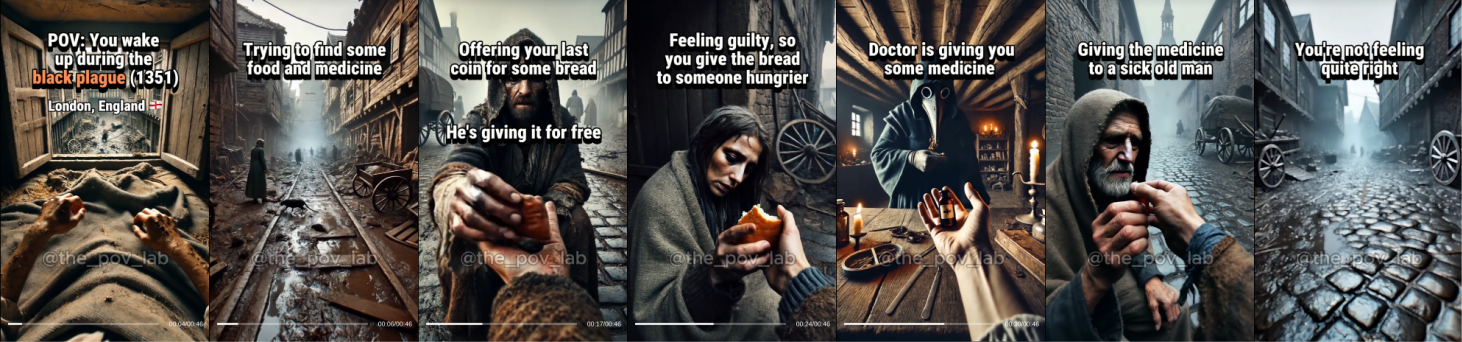}
  \caption{Sequence of scenes from a history POV TikTok (\textit{\@the\_pov\_lab}, post from 12/02/2025, URL: \url{https://vm.tiktok.com/ZNRYduURM})}
  \label{fig:fig1}
\end{figure}
\\
In spring 2025, a viral trend on TikTok popularized AI-generated, animated point-of-view (POV) scenes depicting historical events. \autoref{fig:fig1} shows a frame sequence from one such video set during the Black Death, which accumulated 4.4 million likes and approximately 21,000 comments.\footnote{As of December 16, 2025}
TikTok has become one of the most widely used digital spaces for young people, with particularly high adoption among 18–24-year-olds \cite{Bobzien2025-jw}. The platform also functions as an information source: 67.5\% of 18–39-year-olds report that digital formats are (very) important to them when seeking information about National Socialism in Germany \cite{landecker2024digitaleerinnerung}. TikTok's success is due to both its short-video format and its recommendation algorithm that drives the central \textit{ForYouPage} through a balance of \textit{exploitation} (reinforcing existing interests) and \textit{exploration} (introducing new topics) \cite{Neubert2024-pr, Vombatkere2024-kp}. The algorithm also shapes creator behavior, incentivizing platform-specific practices such as layering videos with text, images, and audio, producing a characteristically dense, fast-paced storytelling style \cite{Neubert2024-pr}, as seen with the history POVs. 
\begin{figure}[ht]
  \centering
  \includegraphics[width=0.8\textwidth]{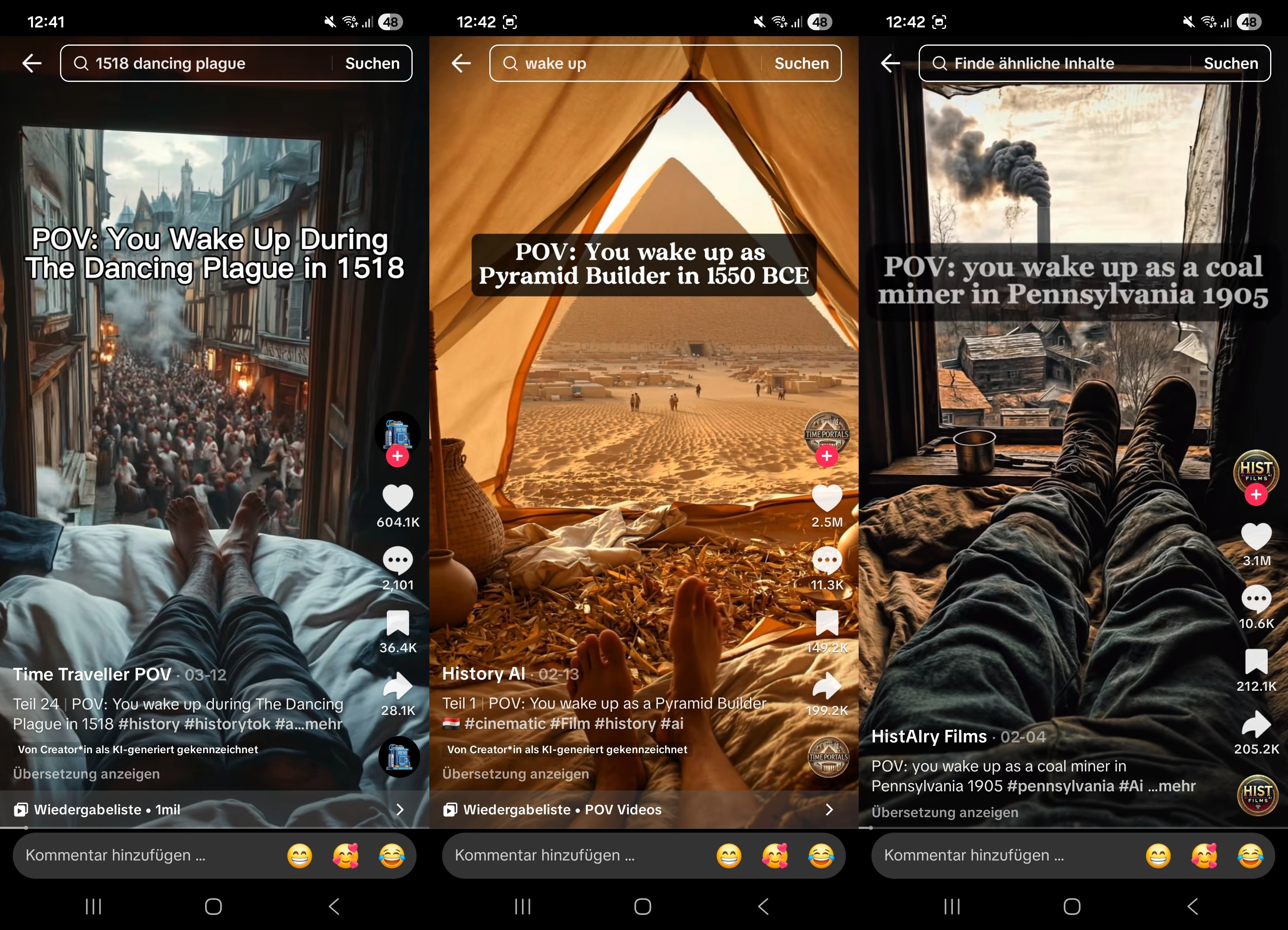}
  \caption{Various history POV TikToks(\textit{\@timetravellerpov}, post from 12/03/2025, URL: \url{https://vm.tiktok.com/ZNR8M73Hr}, \textit{\@timeportals}, post from 13/02/2025, URL: \url{https://vm.tiktok.com/ZNR8reuet}, \textit{\@histairy\_films}, post from 04/02/2025, URL: \url{https://vm.tiktok.com/ZNR8rJnL9})}
  \label{fig:fig2}
\end{figure}
\\
These TikToks follow a consistent narrative template: short, first-person scenes depicting a (fictionalized) historical figure in a historical setting, typically opening with the formula "POV: You wake up…" (see \autoref{fig:fig2}). Visual immersion is achieved through on-screen hands or body parts, text overlays, and atmospheric soundscapes, resulting in a condensed narrative format that focuses less on historical contextualization and more on immediate \textit{experience}. 
\\
Not much is known about the creators of these videos – only a BBC article and several web magazine pieces provide some insight \cite{rufo2025aihistory, uptonclark2025aitiktokhistory}. Creators like \textit{\@the\_pov\_lab} and \textit{\@timetravellerpov} describe a creation process of 4 to 8 hours: historical "research" via ChatGPT, followed by the generation of imagery and its animation using various AI tools. Regarding the videos' potential for disinformation, the creators argue that the videos are primarily designed to spark curiosity and encourage independent research, not to serve as historical sources. 
\\
To gain further insight into the kind of content shared during the history POV trend, as well as its reception, we initially created a small dataset using a \textit{copy, paste, and process} strategy as part of a preliminary study \cite{Brolich2026-py}. 
The study presented in this paper is broader in scope and pursues three objectives:
\begin{enumerate}
    \item to map the thematic landscape of the trend, 
    \item to examine whether videos on different historical topics (namely, the Black Death and the Holocaust as two in-depth case studies) elicit different comment responses, and
    \item to methodologically evaluate the viability of a simple exploratory data collection approach, as seen in the preliminary study, in contrast to an approach using the official TikTok Research API\footnote{\url{https://developers.tiktok.com/products/research-api} [Accessed: \today]}.
\end{enumerate}
After an overview of related work in \autoref{sec:related-work}, we describe our methodology in \autoref{sec:methodology}, present our results in \autoref{sec:results} and conclude with a brief discussion and outlook in \autoref{sec:conclusion}.

\section{Related Work}
\label{sec:related-work}
TikTok is of growing interest to historians both as a platform for historical education and as a space where diverse users actively negotiate and perform historical narratives \cite{oetzel}. Scholarly and public awareness of history-related TikTok content began to develop in response to the controversial \#HolocaustChallenge in summer 2020, in which creators staged themselves as Holocaust victims \cite{Divon2023-qk}. While TikTok research has since expanded rapidly, topic-specific case studies remain scarce \cite{Cervi2023-vn}. This is partly due to the platform's fast-moving nature and the technical, legal, and ethical challenges researchers face. Although TikTok provides a Research API, access is restricted and content archiving is not supported \cite{Berg2023-ay}. Robbert-Jan Adriaansen offers an early systematic overview of historical content via the API \cite{Adriaansen2022-jp}. However, most studies of history-related TikTok content rely on \textit{digital ethnography}: a new account is created, and content encountered via the \textit{ForYouPage}, hashtags, or specific accounts is documented as field observations over a defined period of time \cite{Berg2023-ay, Ackermann2025-oz}.

\section{Methodology}
\label{sec:methodology}
Given that our analysis considers both video captions and user reception, we first outline the methodological approach for video caption collection and analysis in \autoref{methodology:topics}, followed by the procedure for comment collection and analysis in \autoref{methodology:comments}.

\subsection{Data and Methodology for the Analysis of Captions}
\label{methodology:topics}
\begin{figure}[ht]
  \centering
  \includegraphics[width=0.8\textwidth]{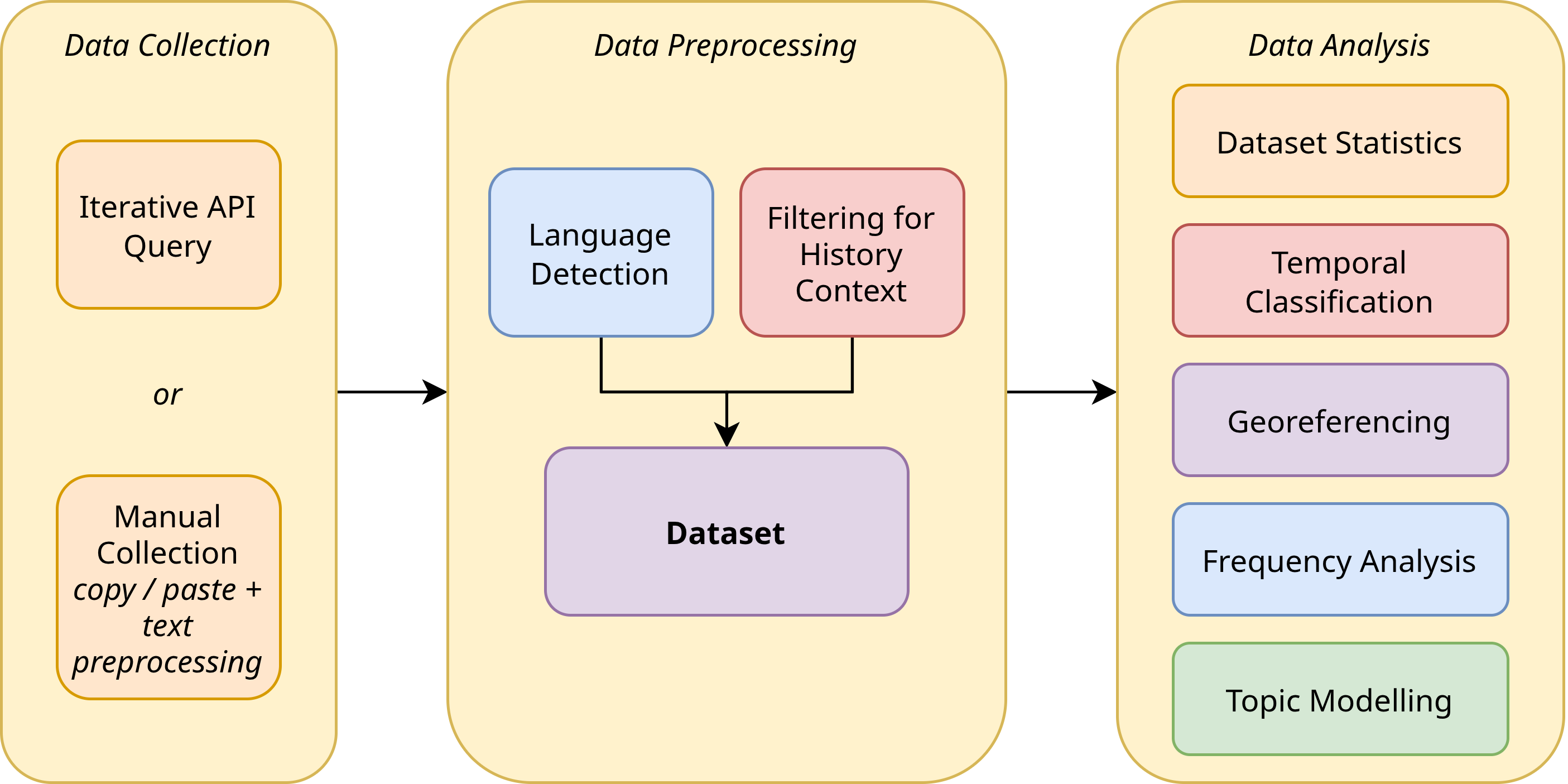}
  \caption{ Methodological approach for identifying key themes in the videos}
  \label{fig:fig3}
\end{figure}
To gain an initial overview of the POV trend, we identified recurring topics of the videos in two steps. In the preliminary study \cite{Brolich2026-py}, a first, smaller dataset of 210 videos was collected with a newly created account and analyzed through spatiotemporal categorization and frequency analysis to identify recurring themes. To systematically validate these findings and to establish a broader empirical basis, a second step involved accessing the TikTok Research API\footnote{To access the API, researchers must apply with a project proposal. The application form does not only ask for detailed information on the researcher's background and their funding, but also requires them to provide a project summary, their research design, research hypotheses, a literature review, their expected outputs, and a data protection plan. In our case, TikTok gave their approval after a few days. 
It should be noted that access to the TikTok Research API is geographically restricted (US, EEA, UK, and Switzerland) and limited to accredited academic institutions, raising concerns about research equity and global accessibility. Additionally, the standard API tier only returns data from users aged 18 and over. Access to data from users under 18 requires approval from the researcher's institutional ethics review board.}. 
For the period from January to September 2025, this yielded a substantially larger dataset consisting of the metadata for 5,565 history POV videos. The methodological approach of the pilot study was extended accordingly, combining spatiotemporal categorization and frequency analysis with topic modeling. \autoref{fig:fig3} illustrates the overall workflow for both studies.
\\
Although the AI-generated visuals constitute the central element of the videos and the primary trigger for user interaction in the form of likes, shares, favorites, and comments, they are not the focus of this analysis. Instead, the study examines topics and comments, as these are where reception, reaction, and evaluation of the content become legible. A further challenge for quantitative analysis is that the videos themselves, unlike captions, cannot be retrieved via the Research API.
\\
The API queries used the search terms \textit{history pov}, \textit{pov you wake up}, 
\textit{pov waking up}, and \textit{pov imagine}, as querying \textit{history pov} through the API
returned considerably fewer relevant results than the platform's native search mechanism.\footnote{Multiple factors contribute to this divergence between API results and in-app search results. The API does not return data from users under 18. The video metadata API also does not query live online data, but a large offline dataset, which takes some time to update. It is also possible that the in-app search algorithm simply produces more 'widespread' results, as its inner workings are essentially a black box. It has also been found that the research API does not provide complete and consistent data access, with a substantial number of videos not being available via the research API for no ascertainable reason \cite{entrenaserrano2025tiktoksresearchapiproblems}.}
Data collection took place retrospectively in December 2025, with the query window being set from January 1 to September 30, 2025, covering the trend's peak engagement in February and March while also allowing us to trace its temporal trajectory.
\\
This broad query strategy yields a large but noisy dataset that requires post-hoc filtering. Due to daily API rate limits\footnote{As of June 2026, the daily limit is set at 1,000 requests per day. The Both the API for video metadata and the API for comments can return 100 records per request.}, data collection was implemented as an iterative process. Unexplained internal server errors meant that data for several query-date combinations could not be retrieved despite repeated attempts. However, given the considerable overlap between the different search terms used, a large proportion of the affected videos was presumably still captured under alternative queries.  Since the video metadata was not all captured on the same date, engagement metrics such as like counts represent point-in-time snapshots; however, as most videos had been in circulation for several months by the time of collection, comparability is assumed to be sufficient. All available metadata fields (creator, like and comment counts, creation date, hashtags, captions) were queried and subsequently stored in JSON format.
\\
In a preprocessing step, captions were enriched with language labels using the Python library \texttt{langdetect}\footnote{\url{https://pypi.org/project/langdetect} [Accessed: \today]}, applied to emoji- and hashtag-stripped caption text to reduce misclassification. Afterward, non-English videos were excluded, as all downstream processing is designed for English-language input, and full multilingual support, using, e.g., automated translation tools, would have entailed disproportionate overhead and constitutes a desideratum for future work.
\\
To identify relevant content, i.e., content relating to history, from the broad query results, videos were retained if the keyword \textit{history} appeared in the caption, hashtags, or account name. While this constraint cannot fully exclude non-historical content (e.g., \textit{POV: You wake up as Athena}), manual filtering was infeasible at this scale. Duplicate entries were also removed. The final filtered dataset comprises 5,565 English-language videos with a history context, reduced from an unfiltered pool of 28,163 API results.
\subsection{Data and Methodology for the Analysis of Comments}
\label{methodology:comments}
\begin{figure}[ht]
  \centering
  \includegraphics[width=0.8\textwidth]{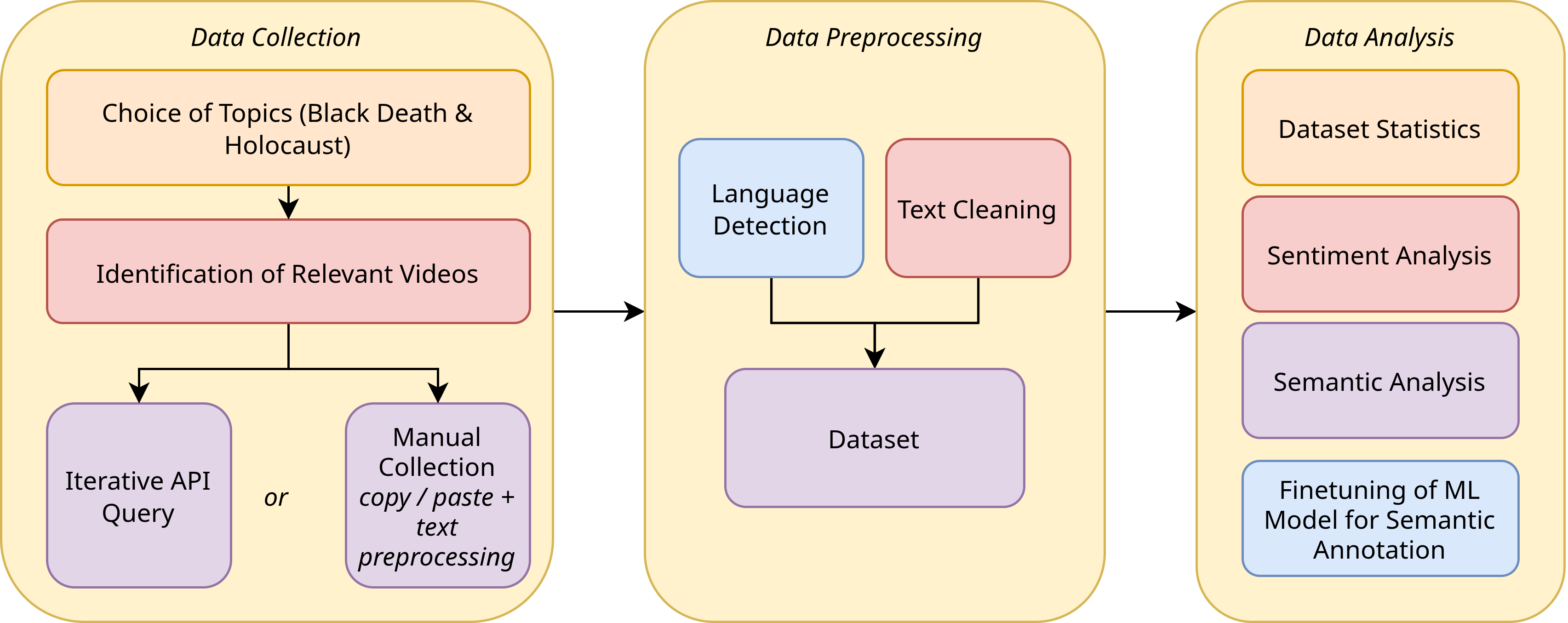}
  \caption{Methodological approach for reception analysis by comparison of two differing topics}
  \label{fig:fig10}
\end{figure}
A further analytical layer examined the reception of the videos. Two topics were selected 
for comparison: the Black Death and the Holocaust. They were chosen primarily for their 
differing degrees of historical sensitivity, in order to assess whether topic choice elicits systematically different comment responses. Black Death videos additionally serve as a 
representative "typical" POV topic: they are among the most numerous in the 
dataset and achieved particularly widespread early engagement.\footnote{In the dataset, a video about the Black Death is, chronologically speaking, the second video to receive
over a million likes and the first to surpass 4 million likes.} Holocaust videos, by contrast, are 
considerably fewer in number and received less widespread engagement, which is 
reflected in the available data volume.
\\
Reception analysis was likewise conducted in two stages: first as part of the exploratory 
pilot study, then on a larger dataset obtained via the TikTok Research API. For each 
topic, relevant videos were identified, comments and their metadata were collected (manually in the 
pilot study, via API in the main study), and two equally sized, English-language datasets 
were compiled. The restriction to English-language comments is again pragmatically 
motivated, primarily by the requirements of the machine learning model. The pilot dataset was subsequently 
annotated semantically, and a classification model was trained and evaluated on the 
annotated data (for methodological details, see \cite{Brolich2026-py}). For the larger dataset, the labeling scheme was refined and extended, 
and sentiment analysis was additionally performed. \autoref{fig:fig10} 
illustrates the overall methodological workflow for the reception analysis.
\\
The API allows comments on specific videos to be queried. Videos were first selected 
from the video metadata dataset by filtering for relevant keywords (e.g., \textit{plague} for the 
Black Death topic; \textit{anne frank}, \textit{jew} for the Holocaust). The considerably larger number of Black Death-related videos is reflected in the volume of comments collected: 16,390 for Black Death videos versus 5,855 for Holocaust videos. 
All available comments were collected in an iterative process due to API quota constraints.\footnote{This is dependent on which comments the API provides access to. As outlined before, the retrieved comment data is incomplete and consistent, due to both constraints the API makes, such as no data from users under 18, and unknown reasons. For example, for the video \emph{POV: You wake up during the black plague (1351)} from TikTok user \emph{the\_pov\_lab}, we were able to collect 14,676 of about 20,700 comments displayed on the app, which corresponds to roughly 70\%.}
\\
Language identification of the comments posed a particular challenge, as social media comments are considerably harder to classify than video captions: they tend to be very short and frequently contain internet slang, emojis, and non-standard or erroneous spellings. Facebook's \texttt{FastText}\footnote{\url{https://fasttext.cc/} [Accessed: \today]} model was therefore used for language detection, as it produces robust results on noisy or unconventionally written text and outputs a confidence score for each prediction. Non-English comments were excluded by retaining only those classified as English with at least 80\% confidence. Emojis were 
converted to their textual representations (e.g.,
\texttt{:crying\_face:}), which is relevant for downstream sentiment analysis. Only top-level comments (i.e., excluding replies) containing at least two words were retained to ensure meaningful semantic classification.\footnote{While reply chains could in principle be reconstructed via their parent comment IDs, their content is often elliptical or referential, making automated classification unreliable without a substantially different analytical approach. Their analysis is therefore left for future work. The two-word threshold reflects a similar trade-off: although single-word comments (e.g., \emph{terrible}) can convey clear sentiment, they more often consist of ambiguous reactions or fillers (e.g., \emph{wow}, \emph{lol}, or emojis) that are difficult to classify reliably at scale, and the threshold was adopted as a pragmatic heuristic to reduce such noise.} Within each video, duplicate comments 
were deduplicated to prevent skewing the machine learning model's training. 
\\
After preprocessing, 1,169 comments remained for the Holocaust videos and 6,747 for 
the Black Death videos. A balanced dataset of 1,000 comments per topic was then created 
via random sampling, as class balance is important for training the classification model.
\section{Analysis and Results}
\label{sec:results}
We compare the results of the analysis of the video captions for both approaches in \autoref{results:topics}, before focusing on the results for the reception analysis in \autoref{results:reception}.

\subsection{Topics of the videos}
\label{results:topics}
Key findings from the preliminary study \cite{Brolich2026-py} regarding topics and time periods show that the videos span past, present, and future settings, with a focus on modern and contemporary events. Dominant themes include military history, prominent historical figures, and emotionally charged or shocking subjects such as disasters and wars, i.e., narrative strategies clearly oriented toward maximizing attention and visibility. This is reflected in the most frequent caption terms, which cluster around executions, final days, and catastrophes.
\begin{figure}[ht]
  \centering
  \includegraphics[width=0.8\textwidth]{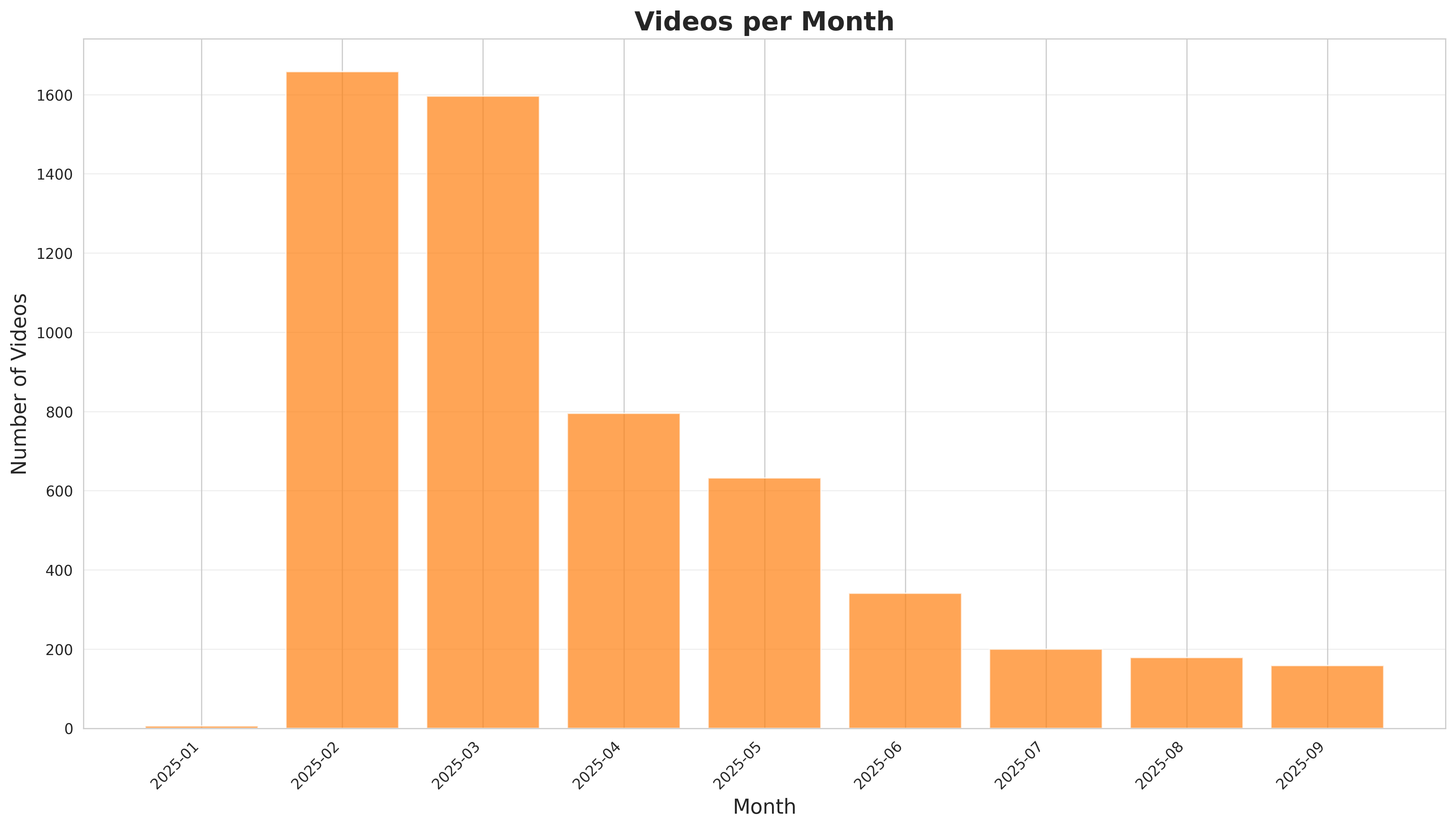}
  \caption{Distribution of video uploads over time}
  \label{fig:fig4}
\end{figure}
\\
The 5,565 videos in the API dataset were uploaded by 2,023 distinct creators from 92 countries, with the majority originating from the United States (28.4\%), the United Kingdom (27\%), and Germany (13.3\%). The temporal distribution of uploads shows that the first history POVs appeared in late January, with the trend gaining momentum in February and peaking across February and March before gradually declining (see \autoref{fig:fig4}). Videos range from a duration of 0 seconds\footnote{These "videos" consist of a slide show of several photos in the history POV format.} to over 7 minutes, averaging 35.9 seconds in length.
\begin{table}[h]
\centering
\begin{tabular}{lrrrrr}
\toprule
 & \textbf{Number of views} & \textbf{Number of likes} & \textbf{Number of comments} & \textbf{Number of shares} \\
\midrule
Average & 137.491  & 9.371      & 57      & 794           \\
Median       & 876      & 38         & 1       & 1            \\
Maximum      & 67.575.927 & 4.370.138 & 21.217 & 357.158  \\
Sum       & 765.135.460 & 52.150.025 & 314.922 & 4.420.479 \\
\bottomrule
\end{tabular}
\caption{Engagement metrics of the dataset}
\label{tab:engagement}
\end{table}
\\
Engagement metrics reveal a highly skewed distribution (see \autoref{tab:engagement}): 
the majority of videos received little to no interaction, while a small number achieved 
substantial reach. Aggregated figures nonetheless confirm the trend's overall scale.
\begin{figure}[ht]
  \centering
  \includegraphics[width=0.8\textwidth]{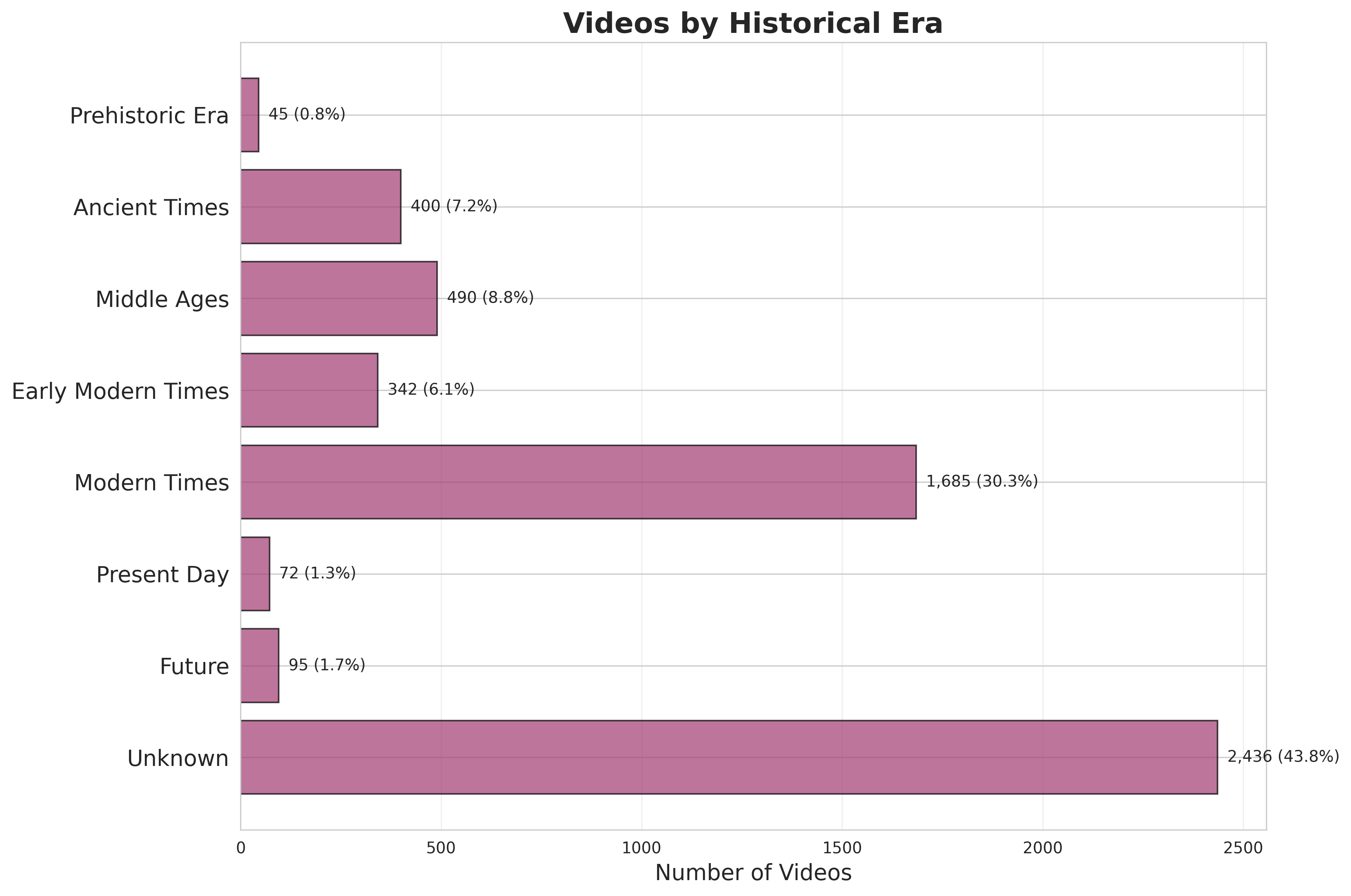}
  \caption{Number of videos per time period}
  \label{fig:fig5}
\end{figure}
\\
To better understand the thematic scope of the dataset, videos were assigned to 
historical epochs based on years or time periods extracted from cleaned captions 
(stripped of hashtags, emojis, and the opening POV formula) using rule-based methods. 
The same time period taxonomy as in the pilot study was applied: years before 3100 BCE were classified as \textit{Prehistoric Era}; years from 3100 BCE up to and including 476 CE as \textit{Ancient Times}; years from 477 to 1500 CE as \textit{Middle Ages}; years from 1500 to 1749 CE as \textit{Early Modern Times}; years from 1750 to 2024 CE as \textit{Modern Times}; the year 2025 as \textit{Present Day}; and years after 2025 as \textit{Future}.\footnote{This, of course, raises questions about whether and, if so, which numerical boundaries can be drawn between time periods.}
\\
A substantial portion of videos contained no extractable dates, as seen in \autoref{fig:fig5}. The results visualized in \autoref{fig:fig6} and \autoref{fig:fig7} confirm the pilot study finding that contemporary history dominates, with several events appearing repeatedly (e.g., building the pyramids, the Black Death, the sinking of the Titanic, or the Chernobyl disaster). The larger dataset additionally reveals a wider temporal spread and greater thematic diversity, 
supporting the case for API-based collection over manual approaches.
\begin{figure}[ht]
  \centering
  \includegraphics[width=0.8\textwidth]{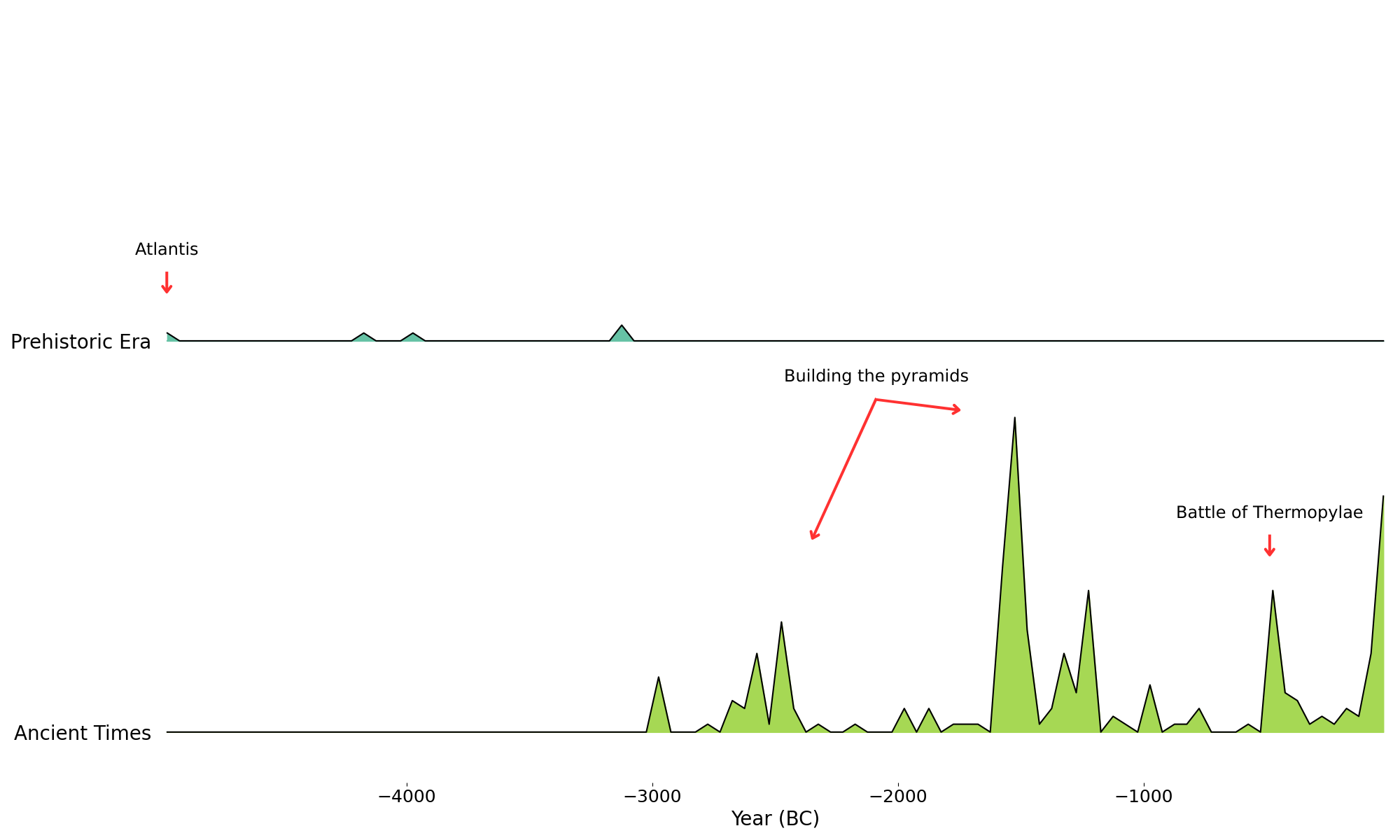}
  \caption{Temporal distribution of the videos B.C.}
  \label{fig:fig6}
\end{figure}
\begin{figure}[ht]
  \centering
  \includegraphics[width=0.8\textwidth]{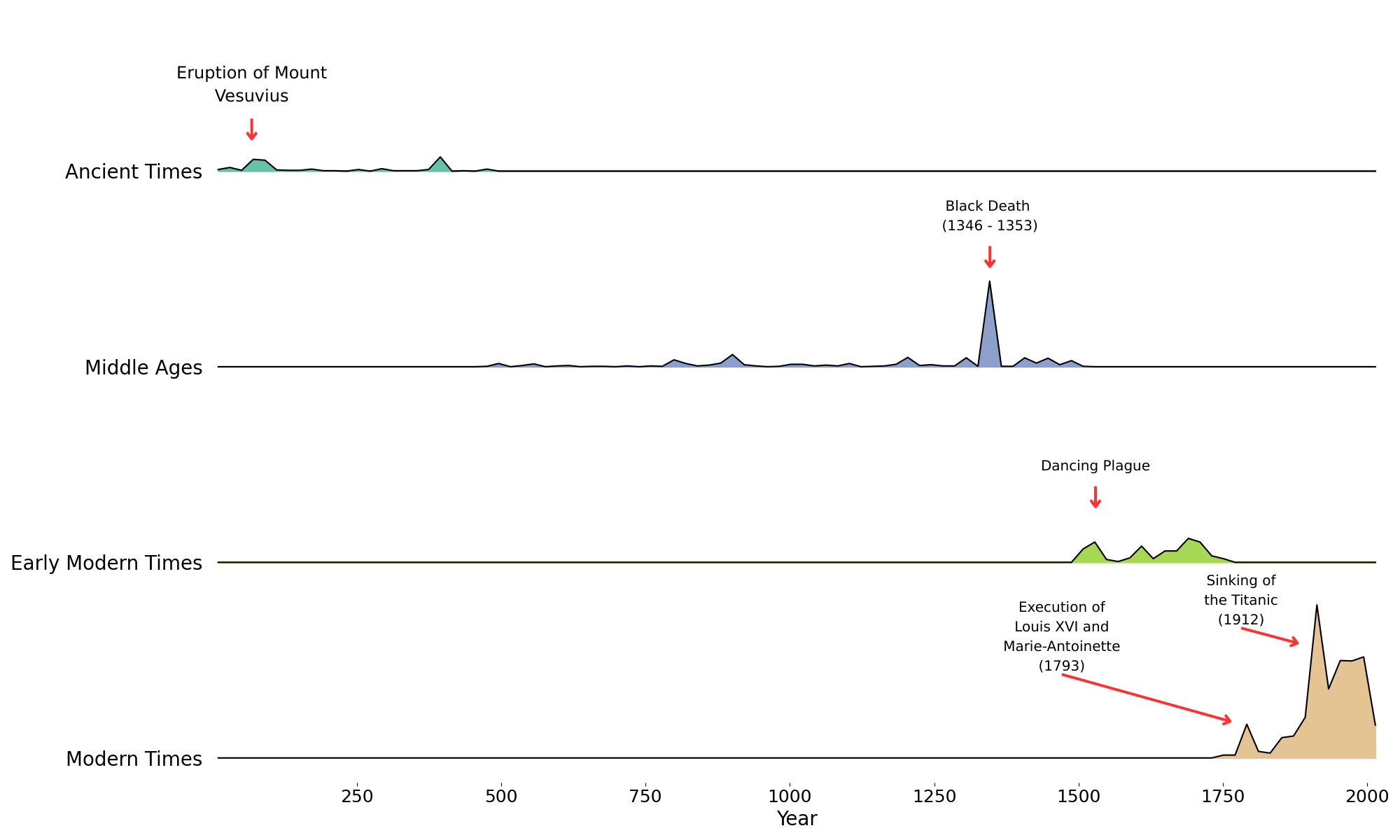}
  \caption{Temporal distribution of the videos A.D.}
  \label{fig:fig7}
\end{figure}
\\
Named entity recognition via \texttt{spaCy}\footnote{\url{https://spacy.io/} [Accessed: \today]}, combined with the lexicon-based libraries 
\texttt{pycountry}\footnote{\url{https://github.com/pycountry/pycountry} [Accessed: \today]} and \texttt{geotext}\footnote{\url{https://github.com/elyase/geotext} [Accessed: \today]}, was used to extract place references from cleaned 
captions. Detected locations were georeferenced using \texttt{Nominatim}\footnote{\url{https://nominatim.org} [Accessed: \today]}. Non-localizable or fictional place names (e.g., \textit{Gotham City}, \textit{Atlantis}) were excluded manually, though a residual error rate cannot be ruled out. The results are visualized in \autoref{fig:fig8}\footnote{They are also accessible via a DARIAH-DE Geo-Browser instance, where they were dated according to their caption or as 2025 if no date was given in the caption: \url{https://geobrowser.de.dariah.eu/index.html?csv1=https://cdstar.de.dariah.eu/dariah/EAEA0-8026-BB8E-6B00-0} [Accessed: \today]}. 
\begin{figure}[ht]
  \centering
  \includegraphics[width=\textwidth]{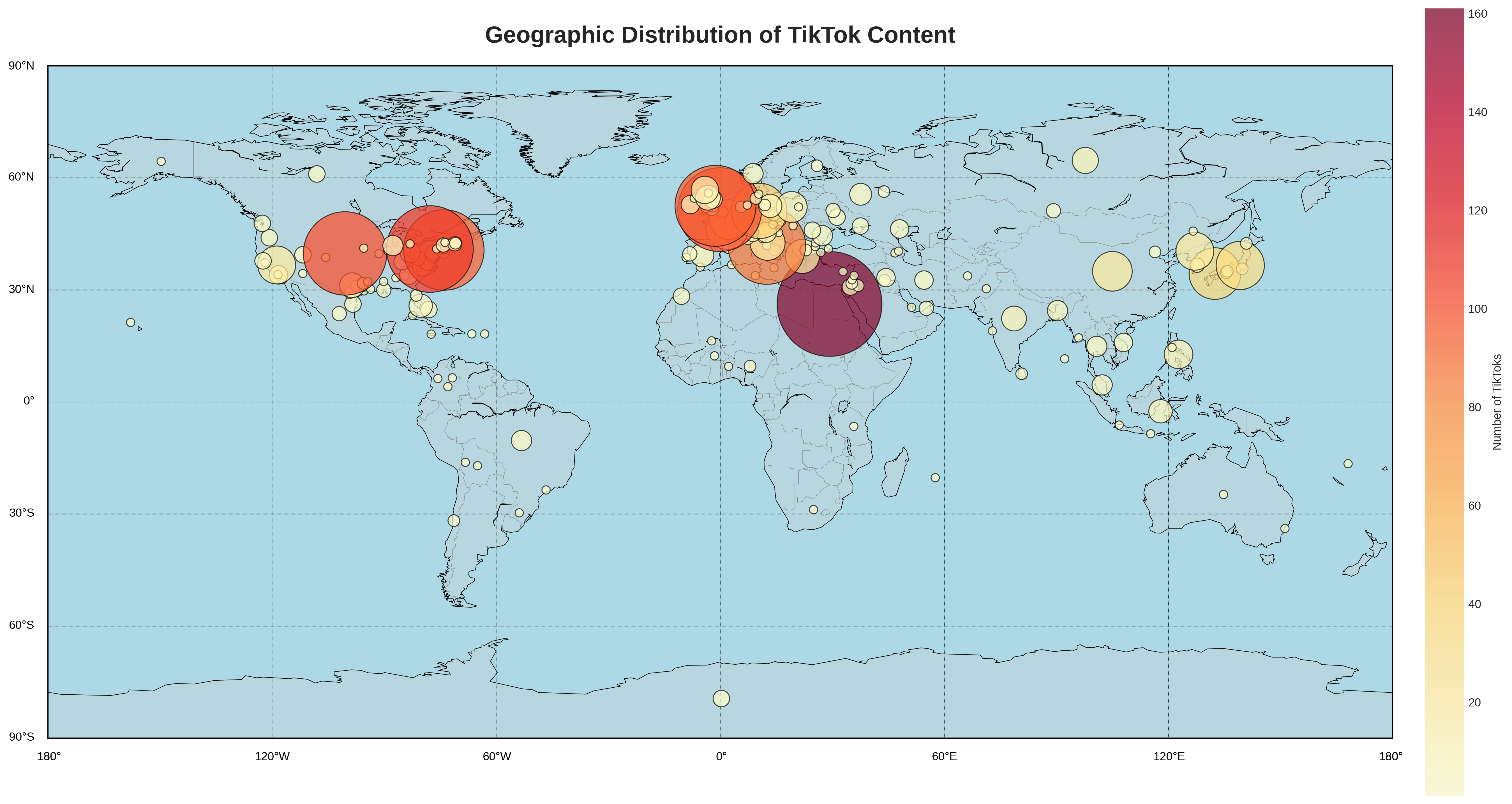}
  \caption{Geographical distribution of identified place names in the video captions}
  \label{fig:fig8}
\end{figure}
The geographic distribution is concentrated on Europe and the United States, reflecting videos' production origins; multilingual API queries could potentially 
broaden this coverage. Even within this limitation, the geographic scope of the data once again demonstrates the added value of API access over manual collection methods.
\\
Bigram frequency analysis (see \autoref{fig:fig9}) of cleaned captions confirms the pilot study finding that generally popular and
emotionally charged topics such as disasters, wars, and catastrophes dominate. Compared 
to the preliminary study, the focus on individual historical figures (e.g., Marie Antoinette, 
Cleopatra) is less pronounced, while everyday historical scenarios across different decades 
(e.g., \textit{teenager 1960s}, \textit{teenager 80s}) also appear as a recurring theme.
\begin{figure}[ht]
  \centering
  \includegraphics[width=0.8\textwidth]{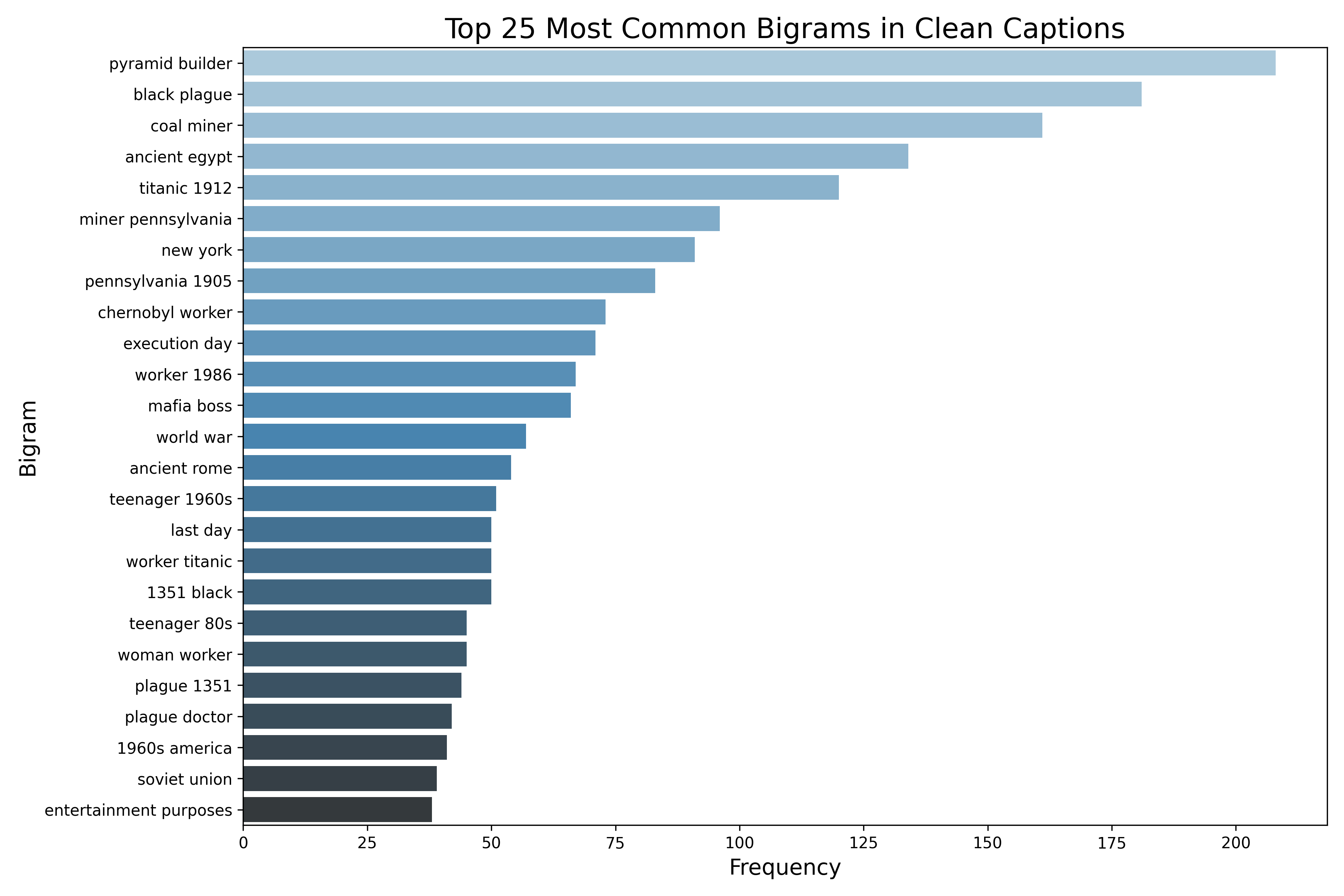}
  \caption{Most common bigrams in the captions}
  \label{fig:fig9}
\end{figure}
\\
Following Robbert-Jan Adriaansen \cite{Adriaansen2022-jp}, we attempted to combine cleaned captions with in-video text overlays for \texttt{BERTopic}\footnote{\url{https://maartengr.github.io/BERTopic/index.html} [Accessed: \today]}-based topic modeling. However, only a small fraction of videos made use of TikTok's native text overlay feature. Additional on-screen text was presumably added via external editing tools and was thus not accessible via the API. A combined caption and overlay corpus of sufficient size could therefore not be assembled, as manual transcription of overlay text was not feasible given the scope of this study. The resulting topics were consequently semantically diffuse, difficult to interpret, and often highly similar to one another, likely a consequence of caption brevity\footnote{As of June 2026, TikTok captions can consist of a maximum of 4,000 characters.}. For this dataset, rule-based spatiotemporal classification and frequency analysis proved more informative 
than topic modeling.

\subsection{Reception of the videos}
\label{results:reception}
\begin{figure}[ht]
  \centering
  \includegraphics[width=\textwidth]{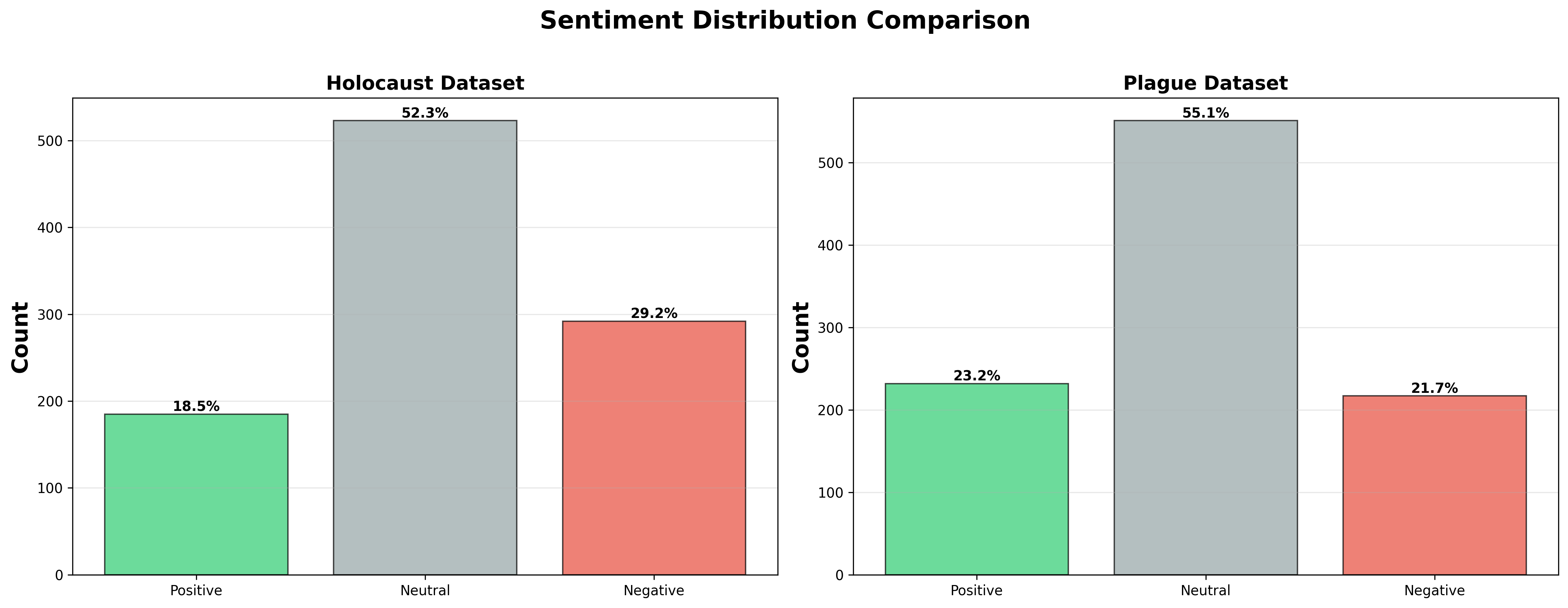}
  \caption{Number of positive, neutral and negative comments in both datasets}
  \label{fig:fig11}
\end{figure}

\begin{figure}[ht]
  \centering
  \includegraphics[width=\textwidth]{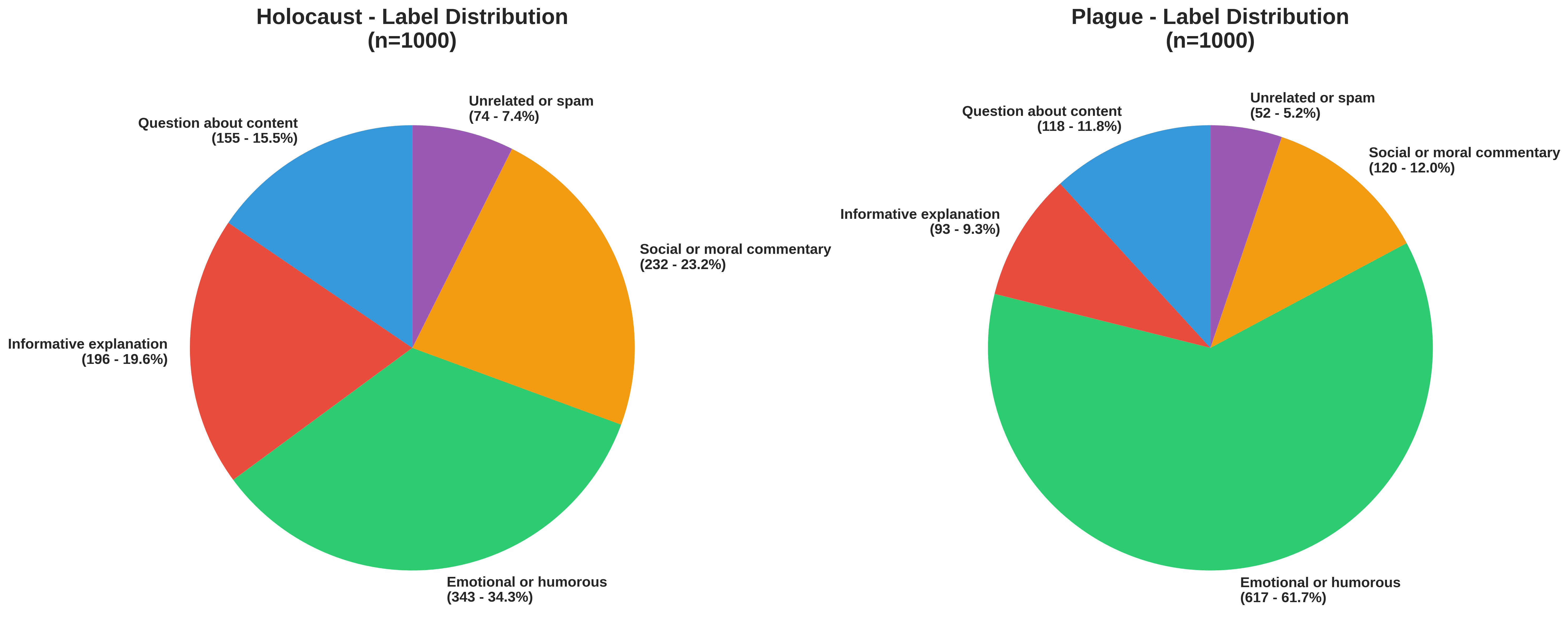}
  \caption{Distribution of labels in both datasets}
  \label{fig:fig12}
\end{figure}

\begin{figure}[ht]
  \centering
  \includegraphics[width=\textwidth]{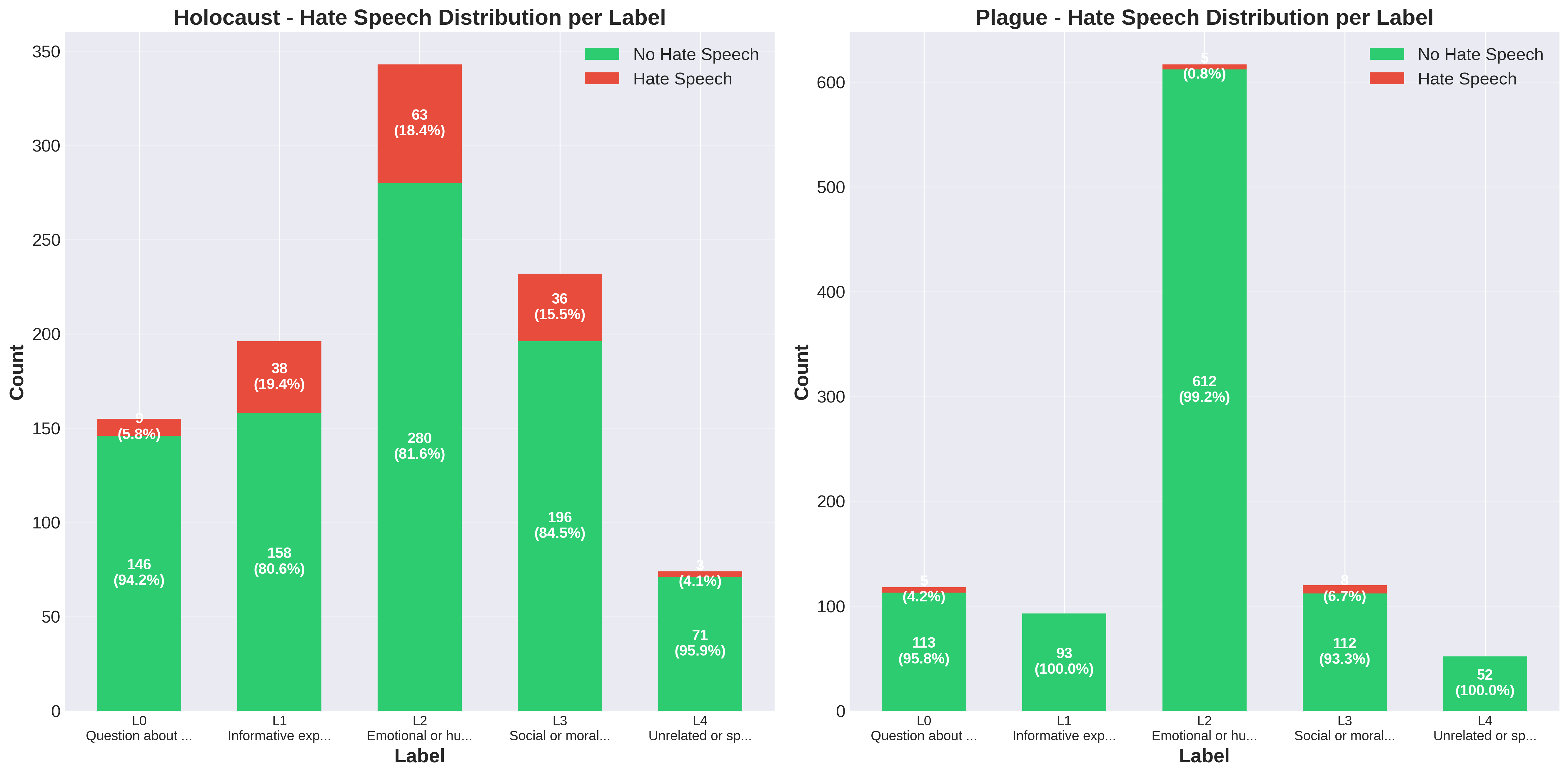}
  \caption{Distribution of labels in both datasets}
  \label{fig:fig13}
\end{figure}

To obtain a baseline understanding of comment sentiment across both topics, a lexicon-based sentiment analysis was conducted using \texttt{VADER}\footnote{VADER (Valence Aware Dictionary and sEntiment Reasoner), \url{https://github.com/cjhutto/vaderSentiment} [Accessed: \today]}. As a lexicon-based method, \texttt{VADER} classifies sentiment based on the prevalence of words with positive or negative connotations rather than contextual or nuanced understanding, meaning results indicate only a general tendency. Results (\autoref{fig:fig11}) show that Holocaust-related videos attracted fewer positive and neutral comments but a higher proportion of negative ones.
\\
To better understand the source of these differences, the dataset was subsequently annotated using a multi-dimensional labeling scheme, developed from the pilot study system\footnote{The labeling system used in the pilot study consisted of the following semantic labels: questions about the video's content; emotional, humorous, social commentary, or personal anecdotes; AI-related comments or references to the POV-TikTok trend; disinformation, hate speech; ethical or moral concerns about the video; historical explanations, corrections, or factual context.} and refined through the data itself (see \autoref{tab:labeling}). Each comment was assigned one out of five primary labels: \texttt{question\_about\_content} (0), \texttt{informative\_explanation} (1),  \texttt{emotional\_or\_humorous} (2), \texttt{social\_or\_moral\_commentary} (3), \texttt{unrelated\_or\_spam} (4). Moreover, each comment was tagged with a binary flag~A indicating AI or trend reference, and an analogous 
flag~B for hate speech or disinformation. This structure reflects the empirical reality 
that a comment may simultaneously express humor and reference the POV trend, for 
instance. 
\\Annotation of all 2,000 comments was performed manually by a single 
annotator, using a structured spreadsheet rather than a dedicated annotation tool, as this approach was simple and fully sufficient for the task at hand. The annotator had also annotated the data during the pilot study and is also one of the authors of this paper. While close familiarity with the material was necessary to perform the annotation reliably, this also introduces potential bias into the classification process. To establish reliability, a subsample of the annotated comments was independently reviewed by the second author. Annotation was a challenging task: comments are very short, 
employ platform-specific language, reference other trends or memes, and are frequently 
ambiguous. Nevertheless, the labels are expected to reflect recurring semantic patterns.
\begin{table}[ht]
\centering
\begin{tabular}{llp{6cm}}
\toprule
\textbf{Dimension} & \textbf{Category} & \textbf{Description} \\
\midrule
Label 
  & \texttt{question\_about\_content} (0) & Comment asks about the video content, context, or meaning \\
\addlinespace
  & \texttt{informative\_explanation} (1) & Provides information, historical context, corrections, or clarification (even if factually incorrect) \\
\addlinespace
  & \texttt{emotional\_or\_humorous} (2) & Expresses personal emotion, humor, irony, or anecdotes \\
\addlinespace
  & \texttt{social\_or\_moral\_commentary} (3) & Expresses social observation, critique, ethical or moral judgment \\
\addlinespace
  & \texttt{unrelated\_or\_spam} (4) & Unrelated to the video's content; does not fall into any other category \\
\midrule
Flag A & \texttt{ai\_or\_trend\_related} (0/1) & Comment references AI, POV edits, or TikTok trends \\
\addlinespace
Flag B & \texttt{hate\_speech\_disinformation} (0/1) & Comment contains offensive, hateful, or misleading content \\
\bottomrule
\end{tabular}
\caption{Annotation scheme for comment classification.}
\label{tab:labeling}
\end{table}
\\
As in the preliminary study, the label distribution (\autoref{fig:fig12}) shows that 
Holocaust-related comments contain considerably more socially critical or moral 
statements as well as more explanatory or contextualizing content. Contrary to the pilot 
study, however, comments referencing AI or the POV trend (e.g., \textit{``I was just 
Cleopatra''}) are distributed evenly across both topics (378 vs.\ 366 comments). 
Comments containing hate speech, disinformation, or offensive and discriminatory 
language are markedly more frequent under Holocaust videos (149 vs.\ 18 comments). 
\autoref{fig:fig13} further illustrates that disinformation frequently takes the 
form of ostensibly factual explanation (label~1; e.g., revisionist claims about Red Cross 
statistics\footnote{This reflects a recurring conspiracy theory; 
see \url{https://arolsen-archives.org/en/news/fact-check-this-document-does-not-relativize-the-holocaust} [Accessed: \today].}), 
that inappropriate comments and "jokes" about the Holocaust appear under label~2 (e.g., 
\textit{``Anne Frank loves the showers''}), and that some ostensibly socially critical 
comments contain hate speech (label~3; e.g., \textit{``Imagine being forced to work like 
every other person in society it's normal''} under an Auschwitz video).
\\
The fully annotated dataset was split randomly into 70\% training, 15\% validation, and 
15\% test data. As in the pilot study, \texttt{DistilBERT}\footnote{\url{https://huggingface.co/docs/transformers/model_doc/distilbert} [Accessed: \today]} was selected as the base model, and several fine-tuned instances were trained. One model was trained on the full combined dataset of 2,000 comments; two further models were trained for a specific topic, i.e., on 1,000 comments from a single topic, capturing clearer semantic coherence at the cost of reduced data volume. An additional model was trained with a balanced sampling strategy, ensuring equal representation of all labels and flags during training.
\begin{table}[ht]
\centering
\begin{tabular}{lrrrr}
\toprule
 & \textbf{Mixed dataset} & \textbf{Holocaust-only} & \textbf{Black Death-only} & \textbf{Balanced (mixed)} \\
\midrule
Accuracy: Labels          & 80.33\% & 78.87\% & 81.33\% & 78.00\% \\
F1-Score: Labels          & 77.31\% & 76.57\% & 71.89\% & 74.00\% \\
Accuracy: AI Flag         & 87.00\% & 89.33\% & 82.00\% & 89.33\% \\
F1-Score: AI Flag         & 77.97\% & 84.91\% & 72.16\% & 84.47\% \\
Accuracy: Hate Speech Flag & 94.00\% & 90.00\% & 100.00\% & 94.67\% \\
F1-Score: Hate Speech Flag & 52.63\% & 63.41\% & 100.00\% & 63.64\% \\
\bottomrule
\end{tabular}
\caption{Fine-tuning results for all \texttt{DistilBERT} model variants.}
\label{tab:finetuning}
\end{table}
\\
Results (\autoref{tab:finetuning}) show no single approach to be clearly superior. All 
models achieve moderate but reasonable performance, reflecting the inherent difficulty 
of the task. For imbalanced datasets, the F1-score, which balances precision and 
recall, is a more informative metric than accuracy alone. The topic-specific models 
achieve partially high scores, though this is partly attributable to the smaller dataset 
size.\footnote{In the Black Death subset, for instance, correctly classifying a single 
comment in a rare category such as hate speech is sufficient to achieve 100\% accuracy 
on that label.} Classification of the five primary label categories appears to benefit from 
the greater semantic variation present in the combined dataset.
\\
Overall, none of the models performs the annotation task with satisfying accuracy, but all are 
suitable for meaningfully supporting manual annotation workflows. Given that 
promising results are already achieved at this data scale, model performance could 
likely be improved substantially by expanding the dataset, for instance, through 
targeted API collection for individual topics (particularly the Black Death, for which 
far more comments are available) or by incorporating semantically related topics (e.g., 
\textit{``POV: you wake up during the dancing plague''}). Within the scope of this study, 
however, such expansion was not feasible due to the limited data available for one of 
the two topics.

\section{Conclusion}
\label{sec:conclusion}
The analysis of video topics shows a clear dominance of early modern and contemporary 
history. Emotionally charged subjects such as wars, disasters, and catastrophes serve as 
the primary narrative starting points. Historical inaccuracies are already visible at the caption level, as 
illustrated by the frequent use of ``Black Plague'' instead of ``Black Death'', or a total of 17 videos captioned \textit{``POV: You wake up as a caveman in 40 BC''}.
\\
The reception analysis confirms that topic choice influences comment 
behavior. Holocaust-related videos attract more critical and evaluative responses as well 
as more contextualizing information, but also substantially more hate speech and 
disinformation. Across both topics, AI generation itself is frequently addressed. The hashtag \texttt{\#ai}, for example, appears in approximately 52\% of videos. This suggests that 
the novelty of the technology, combined with sensationalist content, is a key driver of 
the trend's reach and engagement.
\\
Methodologically, the preliminary study demonstrates that meaningful results can be obtained even with compact methods and a small dataset. API access substantially broadens the empirical basis but requires programming expertise and raises concerns regarding 
accessibility, data protection, and completeness: content from users under 18 is 
excluded, and issues of API reliability and documentation were encountered. 
Crucially, the API provides no access to the videos themselves, preventing direct computer-driven 
analysis of the AI-generated imagery, which constitutes a significant limitation for assessing how historical content is visually distorted. Nonetheless, the API-based approach 
largely confirms the pilot study's findings while enabling systematic, large-scale 
analysis of thematic patterns and reception dynamics.
\\
Limitations of the machine learning experiments stemmed primarily from the small and 
thematically constrained dataset, compounded by the semantic ambiguity of TikTok comments. Future work could leverage the API approach and open source translation technology for multilingual analysis, particularly to examine whether non-English content deviates from the Eurocentrism 
observed here \cite{brolich2026}. More broadly, the analytical approaches developed in this study are 
transferable to other history-related content on TikTok and similar platforms.

\bibliographystyle{unsrt}  
\bibliography{references}

@INCOLLECTION{Neubert2024-pr,
  title     = "{Gatekeeper} zum »{Markt} der {Erinnerung}«?",
  booktitle = "Public History -- Angewandte Geschichte",
  author    = "Neubert, Anja",
  publisher = "transcript Verlag",
  pages     = "131--164",
  month     =  dec,
  year      =  2024,
  address   = "Bielefeld, Germany",
  copyright = "http://creativecommons.org/licenses/by-nc-nd/4.0",
  language  = "de"
}

@INPROCEEDINGS{Vombatkere2024-kp,
  title      = "{TikTok} and the art of personalization: {Investigating}
                exploration and exploitation on social media feeds",
  booktitle  = "Proceedings of the {ACM} Web Conference 2024",
  author     = "Vombatkere, Karan and Mousavi, Sepehr and Zannettou, Savvas and
                Roesner, Franziska and Gummadi, Krishna P",
  publisher  = "ACM",
  volume     =  10,
  pages      = "3789--3797",
  month      =  may,
  year       =  2024,
  address    = "New York, NY, USA",
  copyright  = "https://creativecommons.org/licenses/by/4.0/",
  conference = "WWW '24: The ACM Web Conference 2024",
  location   = "Singapore Singapore"
}

@online{rufo2025aihistory,
  author       = {Rufo, Yasmin},
  title        = {‘{Amateur} and dangerous’: {Historians} weigh in on viral {AI} history videos},
  year         = {2025},
  date         = {2025-02-23},
  url          = {https://www.bbc.com/news/articles/cy87076pdw3o},
  note         = {Accessed: 2026-05-06},
  organization = {BBC News}
}

@online{uptonclark2025aitiktokhistory,
  author       = {Upton-Clark, Eve},
  title        = {{This} {AI} trend lets {TikTok} users relive history’s best—and worst—moments},
  year         = {2025},
  date         = {2025-02-19},
  url          = {https://www.fastcompany.com/91280871/this-ai-trend-lets-tiktok-users-relive-historys-best-and-worst-moments},
  organization = {Fast Company},
  note         = {Accessed: 2026-05-06}
}

@MISC{Brolich2026-py,
  title     = "``{I} was {JUST} {Cleopatra} man'' -- ein quantitativer {Zugang} zu
               {KI-generierten} {Geschichtsnarrativen} und ihrer {Rezeption} auf
               {TikTok}",
  author    = "Brolich, Nina and Neovesky, Anna",
  publisher = "Zenodo",
  year      =  2026
}

@ARTICLE{Bobzien2025-jw,
  title     = "{Visualizing} age-specific digital platform usage in {Germany}",
  author    = "Bobzien, Licia and Verwiebe, Roland and Kalleitner, Fabian",
  journal   = "Socius",
  publisher = "SAGE Publications",
  volume    =  11,
  number    =  23780231251319360,
  month     =  aug,
  year      =  2025,
  language  = "en"
}

@online{landecker2024digitaleerinnerung,
  author       = {{Alfred Landecker Foundation}},
  title        = {{Umfrage zur digitalen Erinnerungskultur: Wie informieren sich Menschen in Deutschland über die Geschichte des Nationalsozialismus?}},
  date         = {2024-09-27},
  year         = {2024},
  url          = {https://www.alfredlandecker.org/de/article/umfrage-zur-digitalen-erinnerungskultur-wie-informieren-sich-menschen-in-deutschland-%C3%BCber-die-geschichte-des-nationalsozialismus},
  organization = {Alfred Landecker Foundation},
  note         = {Accessed: 2026-05-06}
}

@inbook{oetzel,
author = {Lena Oetzel}, 
title = {{Geschichte} und {Geschichtswissenschaft} in {Social Media}},
booktitle = {Digital Humanities in den Geschichtswissenschaften},
chapter = {},
address = {Stuttgart, Deutschland},
year = {2023},
pages = {398-413},
URL = {https://www.utb.de/doi/abs/10.36198/9783838561165-398-413},
eprint = {https://www.utb.de/doi/pdf/10.36198/9783838561165-398-413}
}

@ARTICLE{Divon2023-qk,
  title     = "{Performing} death and trauma? {Participatory} mem(e)ory and the
               {Holocaust} in {TikTok} \#povchallenges",
  author    = "Divon, Tom and Ebbrecht-Hartmann, Tobia",
  journal   = "AoIR Selected Papers of Internet Research",
  publisher = "University of Illinois Libraries",
  month     =  mar,
  year      =  2023
}

@ARTICLE{Cervi2023-vn,
  title     = "{TikTok} and political communication: The latest frontier of
               politainment? A case study",
  author    = "Cervi, Laura and Tejedor, Santiago and Blesa, Fernando
               Garc{\'\i}a",
  journal   = "Media Commun.",
  publisher = "Cogitatio",
  volume    =  11,
  number    =  2,
  month     =  apr,
  year      =  2023
}

@INCOLLECTION{Berg2023-ay,
  title     = "{\#InstaHistory} -- {Akteur:innen} und {Praktiken} des {Doing History}
               in den sozialen {Medien}",
  booktitle = "Praktiken der Geschichtsschreibung",
  author    = "Berg, Mia and Lorenz, Andrea",
  publisher = "transcript Verlag",
  pages     = "69--88",
  month     =  dec,
  year      =  2023
}

@INCOLLECTION{Adriaansen2022-jp,
  title     = "{Historical} analogies and historical consciousness:
               User-generated history lessons on {TikTok}",
  booktitle = "History Education in the Digital Age",
  author    = "Adriaansen, Robbert-Jan",
  publisher = "Springer International Publishing",
  pages     = "43--62",
  year      =  2022,
  address   = "Cham",
  language  = "en"
}

@INCOLLECTION{Ackermann2025-oz,
  title     = "{Von Freund*innen lernen: Bildungsinfluencer*innen auf TikTok zwischen Selbstvermarktung und Wissensvermittlung}",
  booktitle = "Digitale Linguistik",
  author    = "Ackermann, Judith",
  publisher = "Springer Berlin Heidelberg",
  pages     = "133--155",
  year      =  2025,
  address   = "Berlin, Heidelberg",
  copyright = "https://creativecommons.org/licenses/by/4.0",
  language  = "de"
}

@misc{entrenaserrano2025tiktoksresearchapiproblems,
      title={{TikTok's Research API: Problems Without Explanations}}, 
      author={Carlos Entrena-Serrano and Martin Degeling and Salvatore Romano and Raziye Buse Çetin},
      year={2025},
      eprint={2506.09746},
      archivePrefix={arXiv},
      primaryClass={cs.CY},
      url={https://arxiv.org/abs/2506.09746}, 
}

@unpublished{brolich2026,
  author    = {Brolich, Nina and Neovesky, Anna},
  title     = {Cross-Cultural (Point of) Views on History on TikTok},
  note      = {Accepted for presentation at DH2026, Daejeon,  July 27 – July 31, 2026},
  year      = {2026}
}

\end{document}